\documentclass[useAMS,usenatbib]{mn2e}
\usepackage{graphicx}

\title[Iron abundances in Barnard\,29 \& ROA\,5701]{Iron abundances of B-type
post-Asymptotic Giant Branch stars in globular clusters: Barnard\,29 in M\,13 and ROA\,5701 in $\omega$\,Cen}

\author[H. M. A. Thompson et al.]
         {H. M. A. Thompson$^{1}$\thanks{email: h.thompson@qub.ac.uk.
   \hspace*{0.3cm}
	 \newline This paper includes data taken at the McDonald Observatory of the University of Texas as Austin, and on the ESO 2.2-m La Silla, programme 0077.D-025(A).}, 
	 F. P. Keenan$^{1}$, P. L. Dufton$^{1}$, R. S. I. Ryans$^{1}$, 
\newauthor	 
	J. V. Smoker$^{1}$, D. L. Lambert$^{2}$ and A. A. Zijlstra$^{3}$
       \\
       $^{1}$Astrophysics Research Centre, School of Mathematics and Physics, Queen's University, Belfast 
BT7 1NN\\
       $^{2}$University of Texas at Austin, Austin, TX 78712--1083, USA
\\
$^{3}$School of Physics and Astronomy, University of Manchester, P.O. Box 88, Manchester M60 1QD}

\date{Accepted  
	Received 
	in original form }

\pagerange{\pageref{firstpage}--\pageref{lastpage}} \pubyear{0000}
 
\def\LaTeX{L\kern-.36em\raise.3ex\hbox{a}\kern-.15em
    T\kern-.1667em\lower.7ex\hbox{E}\kern-.125emX}

\begin{document}

\label{firstpage}

\maketitle

\begin{abstract}
High resolution optical and ultraviolet spectra of two B-type post-Asymptotic Giant Branch (post-AGB) stars in globular
clusters, Barnard\,29 in M\,13 and ROA\,5701 in $\omega$\,Cen, have been analysed using 
model atmosphere techniques. The optical spectra have been obtained
with FEROS on the ESO 2.2-m telescope and the 2d-Coud\'{e} spectrograph on the 2.7-m McDonald telescope, 
while the ultraviolet observations are from the GHRS on the HST. 
Abundances of light elements (C, N, O, Mg, Al and S) plus Fe have been determined from the optical spectra, while the ultraviolet data provide additional Fe abundance estimates from Fe\,{\sc iii} absorption lines in the 1875--1900 \AA\ wavelength region. 
A general metal underabundance relative to young B-type stars is found for both 
Barnard\,29 and ROA\,5701. These results are consistent with the metallicities of the respective clusters, as well as with previous studies of the objects. The derived abundance patterns suggest that the stars have not undergone a gas-dust separation, contrary to previous suggestions, although they may have evolved from the AGB before the onset of the third dredge-up. 
However, the Fe abundances derived from the HST spectra are lower than those 
expected from the metallicities of the respective clusters, by 0.5 dex for Barnard\,29 and 0.8 dex for ROA\,5701. A similar systematic underabundance is also found for other B-type stars in environments of known metallicity, such as the Magellanic Clouds. These results indicate that the 
Fe\,{\sc iii} ultraviolet lines may yield abundance values which are systematically too low by typically 0.6 dex and hence such estimates should be treated with caution. 
\end{abstract}


\section{Introduction}

Post-Asymptotic Giant Branch (post-AGB) stars are short-lived objects (typically 10$^4$ years), of initial mass between 0.8 -- 8 M$_{\sun}$. They have evolved from the AGB, undergone severe mass loss, and will become planetary nebulae \citep{ibe83}. This evolutionary stage has been reviewed by, for example, \citet{kwo93} and \citet{van03}, and is important as it provides an insight into the nucleosynthesis and mixing processes during the later stages of stellar evolution.  
Most post-AGB stars have been identified via their infrared excess, arising from a circumstellar dust shell. This shell
is most readily detected at the earliest stages of post-AGB evolution, when it is relatively hot and close to the star. As a result, 
the majority of post-AGBs found to date have just left the AGB, and are of A, F or G spectral type \citep{oud96}.

A number of abundance studies have been performed for post-AGB stars, with the majority being metal-poor \citep{van03}. 
However, there are many features of post-AGB evolution which are not yet fully understood.
For example, several stars show abundance patterns reminiscent of interstellar
gas, with Si, Ca etc following the very low Fe abundance, while CNO and S are near
solar \citep*{van95}. \citet*{wat92} proposed a mechanism for the formation of such a `cleaned-up' 
photosphere in a post-AGB star, where grain formation occurs in the circumstellar shell 
produced in the AGB phase of the star, and subsequent slow accretion of the gas from this material results in the peculiar abundance
pattern. This is most likely to happen during mass transfer between the components of a binary system, or accretion triggered 
by a binary companion. Indeed, \citet{van95} showed that all the extremely Fe deficient post-AGB  stars known at that time are binaries, providing support for
this hypothesis.

Although most of the known post-AGBs are cool objects, several B-type stars originally classified as being Population\,{\sc i} are now believed to be hot post-AGBs, on the basis of their derived atmospheric parameters and chemical compositions
(see, for example, \citealt{rya03} and references therein). However, although the B-type stars should be evolving from 
the cooler post-AGBs, there are significant differences. The most striking is the 
large C underabundance found for several hot objects, which may imply that these left the AGB before the third dredge-up \citep{rya03}.

These apparent abundance differences between the hot and cool post-AGB candidates
indicate that further investigations are required, so that this
evolutionary stage may be more fully understood. As a first step, it is clearly important to study 
hot post-AGB objects of known initial metallicity, rather than field stars of 
unknown origin. Fortunately, a sample of these is available: the well-known 
UV-Bright stars found in some globular clusters, whose post-AGB nature has been unambiguously established
from their position in the cluster colour-magnitude diagram \citep{lan00}.

A review of UV-Bright stars in globular clusters can be found in \citet{moe01}, while searches for globular cluster post-AGB stars include \citet*{zin72}, \citet*{har83} and \citet{deb85}. Of the B-type post-AGB stars studied to date, only three have well determined 
iron abundance estimates, namely ZNG-1 in M\,10 \citep{moo04}, Barnard\,29 in M\,13 \citep{{moe98},{dix98}} and ROA\,5701 in $\omega$\,Cen \citep{moe98}. \citet{moo04} examined optical spectra of ZNG-1, finding an iron abundance larger than the metallicity of the cluster, M\,10. 
Barnard\,29 and ROA\,5701 have been studied several times, primarily due to their relative brightness, using both optical and ultraviolet spectra (see, for example, \citealt*{{aue74},{nor74},{cac84},{deb85},{con94},{ade94},{dix98},{moe98},{tho06}}). 
However the iron abundances have previously been estimated using only the UV spectra.
In particular, \citet{moe98} have analysed HST observations of these stars and derived Fe abundances which are lower than the cluster metallicities, which they explained by the grain formation process discussed above. 
If Barnard\,29 and ROA\,5701 have such `cleaned-up' photospheres, then one would expect them to 
show similar depletions in Mg and Si, which are also subject to grain formation. However, previous studies of 
Barnard\,29 and ROA\,5701 indicate no depletions in Mg nor Si \citep{{con94},{moe98},{tho06}}.

Clearly, the abundances of Barnard\,29 and ROA\,5701 need revisiting, in particular those for Fe. 
In this paper, high resolution optical and ultraviolet spectra of Barnard\,29 and ROA\,5701 are analysed, using 
model atmosphere techniques to estimate atmospheric parameters and chemical compositions. 
The observations and data reduction are discussed in Section 2, while Section 3 details the analysis methods and results. In Section 4 we present a discussion of our results, and in particular we assess the accuracy of Fe abundance determinations for B-type stars using ultraviolet spectra.

\section{Observations and data reduction}
\label{sec_obs}

Optical spectra of Barnard\,29 and ROA\,5701 were obtained during two observing runs, the former
with the 2.7-m Harlan J. Smith Telescope (McDonald Observatory) on three nights between 2005 April 30 and May 5, and the latter with the MPG/ESO 2.2-m Telescope (La Silla, Chile) on 2006 April 10.
Ultraviolet spectra of both stars were taken from the Hubble Space Telescope (HST) archive. Details of the observations are provided below. 

\subsection{Optical spectra}
\subsubsection{Barnard\,29}

Barnard 29 was observed with the 2d-Coud\'{e} spectrograph \citep{tul95}, using a 2.4 arcsec 
slit to achieve a spectral resolution ($R$=$\lambda$/$\Delta\lambda$) of $\sim$35000 and a wavelength 
coverage of $\sim$3700--10000 \AA. Due to cloud, data were taken only on three nights, resulting in 30 half hour exposures, or 15 hours on-source. A Th-Ar hollow cathode lamp was observed at the beginning and end of each night, with the centres of the lines stable to better than $\sim$0.1 pixel. Test exposures were taken at the start of the run, to ensure that optical ghosts (`the picket fence'; \citealt{che00}) and bad columns did not overlap with wavelength regions of interest, in particular the Fe\,{\sc iii} lines around 4420 \AA. Also, in order to ensure that the
weak absorption line features detected were real, the central wavelength was changed, by $\sim$0.4 \AA, after the first night and remained in the new position for the other two clear nights. 

The data were reduced using standard methods within the Image Reduction and Analysis Facility ({\sc iraf}; \citealt{bar93}). This reduction included bias subtraction and flatfielding, with stellar extraction and cosmic--ray removal performed using {\sc doecslit} \citep{val93} and {\sc scombine}. The signal to noise ratio (S/N) per pixel in the co-added spectrum was $\sim$130 per pixel at $\sim$4420 \AA\
(corresponding to $\sim$180 per resolution element), and this was 
read into the Starlink package {\sc dipso} \citep{how04} for further analysis (see Section {\ref{sec_dip}}).

\subsubsection{ROA\,5701}

The optical spectra of ROA\,5701 were obtained using the high-resolution Fibre-fed Extended Range Optical Spectrograph (FEROS; \citealt{kau99}), with the EV 2K$\times$4K CCD detector and a 79 lines mm$^{-1}$ \'{e}chelle grating. In total, ROA\,5701 was observed for 9 hours, 
resulting in a spectral resolution of $R$ $\sim$46000 and a wavelength coverage of $\sim$3560--9200 \AA. 

The data were reduced on-line, using the standard FEROS Data Reduction System (DRS), implemented with ESO--{\sc midas} software. This reduction includes the extraction and de-blazing of the \'{e}chelle orders, wavelength calibration and merging of the orders. The reduced spectral exposures were combined into a single spectrum using {\sc scombine} within {\sc iraf}, and this was imported into {\sc dipso} for further analysis (see Section {\ref{sec_dip}}). 
An average S/N ratio of $\sim$85 per pixel at $\sim$4420 \AA\ was obtained, corresponding to $\sim$120 per resolution element. 

\subsubsection{Equivalent width measurements}
\label{sec_dip}

\begin{figure}
\includegraphics[angle=270,width=0.5\textwidth]{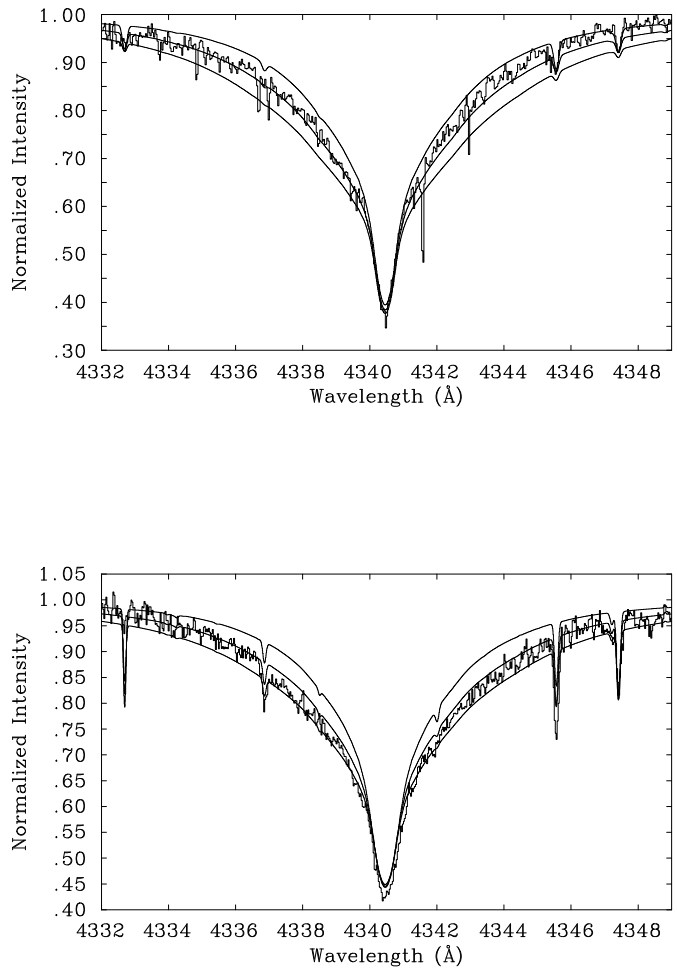}
\caption{Observed and theoretical line profiles for H$\gamma$ at 4341 \AA\ in the spectra
of Barnard\,29 (upper panel) and ROA\,5701 (lower panel). The theoretical profiles have been generated for surface gravities of 2.75, 3.0 and 3.25 dex (upper, middle and lower smooth curves, respectively) for Barnard\,29, implying a surface gravity of 2.95 dex. For ROA\,5701, results have been calculated for surface gravities of 3.0, 3.25 and 3.5 dex (upper, middle and lower smooth curves, respectively), implying log~{\em g} = 3.3 dex.}
\label{fig_hga}
\end{figure}

Initially, the continua of the optical spectra for both Barnard\,29 and ROA\,5701 were normalised using low-order polynomials fitted to spectral regions free from absorption features. The line fitting program {\sc elf} within {\sc dipso} was
then employed to measure absorption line equivalent widths, by fitting Gaussian profiles to the lines (see \citealt{tho06} for further details). 
Balmer lines and diffuse He\,{\sc i} line equivalent widths were not measured, as
instead theoretical profiles were fitted to the normalized data (see, for example, Fig. \ref{fig_hga}). 

\subsection{Ultraviolet spectra}
\label{sec_uv1}

The ultraviolet spectra of Barnard\,29 and ROA\,5701 were obtained directly from the HST Multimission Archive (http://archive.stsci.edu; z3EC0204T and z3EC0104M respectively). They were observed with the Goddard High Resolution Spectrograph (GHRS) using the G200M grating ($R$ = 25000) and the large science aperture (LSR), giving a wavelength coverage of $\sim$1860--1906 \AA. FP splits were not used for these observations \citep{moe98}. 
Barnard\,29 was observed on 1996 November 30 for 4570\,s, and ROA\,5701 on 1996 August 3 for 4711\,s.

Following the standard pipeline reduction, the spectra were co-aligned within {\sc iraf}, using the {\sc stsdas} package (Version 3.4). The combined spectrum for each star was then read into {\sc dipso} and cross-correlated with theoretical spectra to obtain the stellar LSR radial
velocities of --228 and +258 km s$^{-1}$ for Barnard\,29 and ROA\,5701, respectively. 
These are consistent with the radial velocities of M\,13 (--227.2 km s$^{-1}$, \citealt{har96}) and $\omega$\,Cen (+240 km s$^{-1}$, \citealt{nor74}). An interstellar reddening correction, of E(B--V) = 0.02 \citep{har96} for M\,13 and E(B--V) = 0.11 \citep{cac84} for $\omega$\,Cen, was applied to the appropriate spectrum. The spectra were then binned, with a wavelength interval of 0.05 \AA, resulting in a wavelength coverage of 1868--1906 \AA\ for Barnard\,29 and 1865--1903 \AA\ for ROA\,5701.

\section{Data analysis and results}
\label{sec_anar}

\subsection{Model atmosphere calculations}

Non-Local Thermodynamic Equilibrium (non-LTE) model atmosphere grids, calculated using {\sc tlusty} and {\sc synspec} \citep*{{hub88},{hub95},{hub98}}, have been used to derive atmospheric parameters and chemical abundances. A summary of the 
procedures involved are given in \citet{tho06}, while a more detailed discussion of the grids and methods may 
be found in \citet{rya03} and \citet{duf05}, and also at http://star.pst.qub.ac.uk.

Briefly, four grids have been produced with metallicities corresponding to [Fe/H] = 7.5 dex for the Galaxy, 7.2 dex and 6.8 dex for the Large (LMC) and Small (SMC) Magellanic Clouds, respectively, and 6.4 dex for low metallicity regimes. For each metallicity grid, approximately 3000 models were calculated for a range of effective temperatures from 12000 to 35000 K, logarithmic gravities (log~{\em g} = 4.5 dex down to the Eddington limit) and microturbulences of 0 to 30 km s$^{-1}$. The iron abundance was fixed at each grid, while the light element abundances (C, N, O, Mg, Si, S) 
were allowed to vary from --0.8 to +0.8 dex around their base (metallicity) value. 
Iron is assumed to dominate the metal line blanketing. However, the exclusion of nickel in our models should not be a significant source of error in the atmospheric parameters and chemical composition \citep{hub98}. 
The atmospheric structure is therefore defined by the iron abundance (metallicity), effective temperature, gravity and microturbulence. 
It is also assumed that light element abundances can vary with negligible affect on the atmospheric structure, as shown in tests performed by \citet{duf05}. 

Theoretical profiles and equivalent widths of light element transitions were calculated using the models. A GUI interface in IDL allows the user to access the theoretical equivalent widths and interpolate 
these to calculate equivalent widths and abundance estimates for $\sim$200 metal lines, for any set of atmospheric parameters. Abundance increments of 0.4 dex used in the grids were found to be sufficiently
small that no significant errors were incurred due to the interpolation procedures \citep{rya03}.

\subsection{Atmospheric parameters}
\label{sec_at}

The atmospheric parameters of each star were determined from the optical spectra. As the parameters are interrelated, an iterative process is used (see, for example, \citealt*{{kil92},{hun05},{tho06}}). 
Studies of giants in M\,13 yield a cluster metallicity of [Fe/H] = --1.6 \citep*{yon06}, while for $\omega$\,Cen, which contains at least three stellar populations, the main cluster giant population indicates [Fe/H] = --1.6 \citep{ori03}. 
Hence, for both stars the grid corresponding to low metallicity regions (6.4 dex; --1.1 dex lower than Galactic) was used, as this was the lowest metallicity model available and thus the most appropriate based on the metallicities of the respective clusters. 
\citet{lee05} found that using grids of different iron abundances changed the effective temperature estimate by typically $\leq$ 500 K, while the gravity and microturbulence estimates were unaffected, so the use of this grid should not be a major source of error. The adopted atmospheric parameters and associated errors for both stars are listed in Table \ref{tab_ap}.

\begin{table}
\begin{center}
\caption{Adopted atmospheric parameters for Barnard\,29 and ROA\,5701.}
\label{tab_ap}
\begin{tabular}{@{}lccccccc}
\hline
Star		& $T_{\rm eff}$		& log~{\em g}		& $\xi$		      		& $\upsilon$\,sin{\it i}	\\
		& ${\rm 10^{3} K}$	& dex			&  km s$^{-1}$   	&  km s$^{-1}$	\\
\hline\\
Barnard\,29	& 20 $\pm$ 1		& 2.95 $\pm$ 0.1	& 4 $\pm$ 2	        	& 0 			\\
ROA\,5701	& 25 $\pm$ 1		& 3.3 $\pm$ 0.1		& 1 $^{+2}_{-1}$        	& 3		\\      
\hline
\end{tabular}
\end{center}
\end{table}

Effective temperature estimates, $T_{\rm eff}$, were determined using the silicon ionization balance, 
with Si\,{\sc ii}/Si\,{\sc iii} employed for Barnard\,29 and Si\,{\sc iii}/Si\,{\sc iv} for ROA\,5701. 
Surface gravities, log~{\em g}, were derived by over-plotting the observed spectrum in the regions of the Balmer lines with theoretical profiles, with associated errors assessed from the uncertainty in the fitting (Fig. \ref{fig_hga}). 
The microturbulence, $\xi$, was determined using the observed O\,{\sc ii} and Si\,{\sc iii} lines, by removing the dependance of abundance on line strength, i.e. a plot of abundance against line strength having a zero gradient \citep{kil92}.
Individual multiplets were employed
in this procedure, to reduce errors which may arise from using many different multiplets \citep{hun05}.

Initial atmospheric parameters for Barnard\,29 were taken from \citet{con94}, and are $T_{\rm eff}$ = 20000 K, log~{\em g} = 3.0 and $\xi$ = 10 km s$^{-1}$. 
The temperature estimate remained unchanged at 20000 K with an associated error of $\pm$1000 K, the gravity was constrained to log~{\em g} = 2.95 $\pm$ 0.10, 
while a determination of the microturbulence led to a value of $\xi$ = 4 $\pm$ 2 km s$^{-1}$. 

For ROA\,5701, initial parameters were taken from \citet{tho06}, namely $T_{\rm eff}$ = 25000 K, log~{\em g} = 3.25 and $\xi$ = 5 km s$^{-1}$. 
The present analysis led to no change in the effective temperature. As no absorption line was detected for He\,{\sc ii} 4541 \AA, 
assuming a normal helium abundance also allowed us to place an independent upper limit of $T_{\rm eff}$ $<$ 26000 K. 
Therefore, the associated error for $T_{\rm eff}$ is $\pm$1000 K. The gravity was determined to
be log~{\em g} = 3.3 $\pm$ 0.1, and the microturbulence $\xi$ =  1 $^{+2}_{-1}$ km s$^{-1}$.

\subsubsection{Rotational velocity, $\upsilon$\,sin{\it i}}

In addition to the microturbulent broadening, some rotational broadening may occur. The effect of this was investigated by fitting instrumentally broadened non-LTE theoretical profiles for the Si\,{\sc iii} multiplet at 4560 \AA, calculated using the appropriate $T_{\rm eff}$, log~{\em g} and $\xi$ values, to the observed spectra.
The theoretical profiles were rotationally broadened and overlain in order to determine the most appropriate $\upsilon$\,sin{\it i} (see \citealt{tru01} for further details).
For Barnard\,29, the profile fitting resulted in a value of $\upsilon$\,sin{\it i} = 0 km s$^{-1}$, while in the case of ROA\,5701, a $\upsilon$\,sin{\it i} = 3 km s$^{-1}$ was obtained. 

\citet{leh00} discuss the difficulty in differentiating between the effects of the two types of broadening when such low values are obtained. 
In order to place an upper limit on the $\upsilon$\,sin{\it i} values, a microturbulence of 0 km s$^{-1}$ was assumed for both stars, as a non-zero microturbulence would lead to lower $\upsilon$\,sin{\it i} estimates \citep{tru01}. This resulted in, for Barnard\,29 $\upsilon$\,sin{\it i} $\leq$ 6 km s$^{-1}$ and for ROA\,5701, $\upsilon$\,sin{\it i} $\leq$ 3 km s$^{-1}$. 

\subsection{Abundances from optical spectra}

Abundances for both stars were determined using the adopted atmospheric parameters; atomic data were taken from http://star.pst.qub.ac.uk.
Table \ref{tab_all} contains the equivalent width measurements for the observed metal lines, along with the associated abundance estimates. If more than one line contributed to an equivalent width, the lines were treated as blends within {\sc tlusty}.  

There are currently no non-LTE grids available for Al\,{\sc iii}, S\,{\sc iii} or Fe\,{\sc iii}, although sulphur is included in our non-LTE model atmosphere calculations using a relatively simple model. Using {\sc tlusty}, models were calculated, adopting the appropriate atmospheric parameters. These were then employed
to generate theoretical LTE equivalent widths  (see \citealt{hun05}), which were used to derive the abundance estimates shown in Table \ref{tab_all}, with errors calculated by utilising the same method as for other elements. 

\begin{table*}
\begin{minipage}{\textwidth}
\caption{Absorption line equivalent width measurements from optical spectra 
and derived abundances for Barnard\,29 and ROA\,5701.}
\label{tab_all}
\begin{tabular}{@{}llcccccccc}
\hline
\multicolumn{2}{c}{Identification}&\multicolumn{4}{c}{Barnard\,29}&\multicolumn{4}{c}{ROA\,5701}\\
Rest $\lambda$ & Species 	&Observed $\lambda$ & W & Quality$^\dagger$ & Abundance$^\diamond$ &Observed $\lambda$ & W & Quality$^\dagger$ & Abundance$^\diamond$\\ 
(\AA)		&		& (\AA)			& (m\AA)	& & & (\AA)	& (m\AA)	\\
\hline\\

4267.00& C\,{\sc ii}$^a$ 	&$\cdots$ 	&$<$10	 &$\cdots$	& $<$6.13	&$\cdots$ &$<$10	&$\cdots$& $<$6.27  \\	   
6578.05& C\,{\sc ii}	    	&$\cdots$ 	&$<$10	 &$\cdots$	& $<$6.64	&$\cdots$ &$<$10	&$\cdots$& $<$6.54  \\

3995.00& N\,{\sc ii}$^a$ 	& 3995.00 	& 65  	 & a	        & 7.22	        & 3995.00 & 45	       	& a	 & 6.99  \\ 
4035.08& N\,{\sc ii}	 	& 4035.09 	& 12  	 & c	        & 7.62	        &$\cdots$ &$\cdots$     &$\cdots$&$\cdots$\\
4041.31& N\,{\sc ii} 		& 4041.32 	& 25  	 & a	        & 8.30		&$\cdots$ &$\cdots$     &$\cdots$&$\cdots$\\
4043.53& N\,{\sc ii} 		& 4043.52 	& 14  	 & b	        & 7.47	       	&$\cdots$ &$\cdots$     &$\cdots$&$\cdots$\\
4227.74& N\,{\sc ii} 		& 4227.71 	& 8	 & c	        & 7.36	       	&$\cdots$ &$\cdots$     &$\cdots$&$\cdots$\\
4241.78& N\,{\sc ii}$^a$ 	& 4241.79	& 21  	 & a	        & 7.62	       	& 4241.82 & 15	       	& c	 & 7.67  \\ 
4432.73& N\,{\sc ii}$^a$	& 4432.72 	& 12  	 & c	        & 7.39	       	& 4432.75 & 11	       	& c	 & 7.35  \\
4447.03& N\,{\sc ii} 		& 4446.99 	& 30  	 & a	        & 7.28	       	& 4447.03 & 18	       	& b	 & 6.95  \\ 
4601.48& N\,{\sc ii} 		& 4601.47 	& 27  	 & a	        & 7.34	       	&$\cdots$ &$\cdots$     &$\cdots$&$\cdots$\\
4607.16& N\,{\sc ii} 		& 4607.14 	& 21  	 & b	        & 7.29	       	&$\cdots$ &$\cdots$     &$\cdots$&$\cdots$\\
4613.86& N\,{\sc ii} 		& 4613.85 	& 18  	 & a	        & 7.34	       	&$\cdots$ &$\cdots$     &$\cdots$&$\cdots$\\
4621.29& N\,{\sc ii} 		& 4621.38 	& 16  	 & a	        & 7.12	       	&$\cdots$ &$\cdots$     &$\cdots$&$\cdots$\\
4630.54& N\,{\sc ii} 		& 4630.56 	& 50  	 & a	        & 7.28	       	& 4630.55 & 26	       	& b	 & 6.93  \\
4643.09& N\,{\sc ii} 		& 4643.10 	& 27  	 & a	        & 7.33	       	& 4643.10 & 11	       	& c	 & 6.93  \\
5001.10& N\,{\sc ii}$^a$$^b$	& 5001.12 	& 30  	 & b	        & 7.36	       	& 5001.16 & 20	       	& a	 & 6.92  \\
5001.50& N\,{\sc ii}$^a$$^b$	& 5001.47 	& 37  	 & b	        &$\cdots$       & 5001.50 & 25	       	& a	 &$\cdots$\\
5002.70& N\,{\sc ii}$^a$$^b$	& 5002.68 	& 8	 & c	        &$\cdots$       &$\cdots$ &$\cdots$     &$\cdots$&$\cdots$\\
5005.15& N\,{\sc ii} 		& 5005.15 	& 47  	 & b	        & 7.34	       	& 5005.17 & 30	       	& a	 & 6.88  \\
5007.33& N\,{\sc ii} 		& 5007.23 	& 30  	 & c	        & 7.68	       	& 5007.34 & 11	       	& b	 & 6.96  \\

3911.96& O\,{\sc ii}$^a$$^b$	&$\cdots$  	&$\cdots$&$\cdots$      &$\cdots$	& 3911.98 & 27  	& a      & 7.64  \\
3912.09& O\,{\sc ii}$^a$$^b$	&$\cdots$  	&$\cdots$&$\cdots$	&$\cdots$	&$\cdots$ &$\cdots$	&$\cdots$&$\cdots$\\ 
3919.28& O\,{\sc ii} 		&$\cdots$  	&$\cdots$&$\cdots$	&$\cdots$	& 3919.30 & 19  	& b	 & 7.74  \\
3945.04& O\,{\sc ii} 		&$\cdots$  	&$\cdots$&$\cdots$	&$\cdots$	& 3945.04 & 15  	& b	 & 7.62  \\
3954.37& O\,{\sc ii} 		&$\cdots$  	&$\cdots$&$\cdots$	&$\cdots$	& 3954.38 & 29  	& a	 & 7.65  \\
3982.72& O\,{\sc ii} 		&$\cdots$  	&$\cdots$&$\cdots$	&$\cdots$	& 3982.72 & 20  	& b	 & 7.78  \\
4069.62& O\,{\sc ii}$^a$$^b$	& 4069.61 	& 13  	 & b     	& 7.55  	& 4069.64 & 36  	& a      & 7.76  \\
4069.89& O\,{\sc ii}$^a$$^b$	& 4069.87 	& 17  	 & b     	&$\cdots$       & 4069.90 & 47  	& a      &$\cdots$\\
4072.16& O\,{\sc ii} 		& 4072.13 	& 23   	 & a     	& 7.52  	& 4072.17 & 57  	& a      & 7.72  \\
4075.86& O\,{\sc ii} 		& 4075.83 	& 30  	 & a     	& 7.54  	& 4075.87 & 68  	& a      & 7.71  \\
4078.84& O\,{\sc ii} 		&$\cdots$  	&$\cdots$&$\cdots$	&$\cdots$	& 4078.85 & 19  	& b	 & 7.79  \\
4089.29& O\,{\sc ii} 		&$\cdots$  	&$\cdots$&$\cdots$	&$\cdots$	& 4089.29 & 33  	& a	 & 8.02  \\
4132.80& O\,{\sc ii} 		&$\cdots$  	&$\cdots$&$\cdots$	&$\cdots$	& 4132.81 & 24  	& b	 & 7.76  \\
4185.46& O\,{\sc ii} 		&$\cdots$  	&$\cdots$&$\cdots$	&$\cdots$	& 4185.47 & 24  	& a	 & 7.67  \\
4294.79& O\,{\sc ii} 		&$\cdots$  	&$\cdots$&$\cdots$	&$\cdots$	& 4294.77 & 11  	& c	 & 7.56  \\
4303.84& O\,{\sc ii}$^a$	&$\cdots$  	&$\cdots$&$\cdots$	&$\cdots$	& 4303.85 & 15  	& c	 & 7.39  \\
4317.14& O\,{\sc ii}$^a$ 	& 4317.12 	& 10  	 & a     	& 7.35  	& 4317.15 & 37  	& a      & 7.78  \\
4319.63& O\,{\sc ii} 		& 4319.61 	& 13  	 & a     	& 7.52  	& 4319.64 & 46  	& a      & 7.95  \\
4349.43& O\,{\sc ii} 		& 4349.43 	& 25  	 & a     	& 7.59  	& 4349.43 & 71  	& a      & 7.96  \\
4351.26& O\,{\sc ii}$^a$ 	& 4351.25 	& 9   	 & c     	& 7.29  	& 4351.27 & 38  	& a      & 7.54  \\
4366.90& O\,{\sc ii}$^a$ 	& 4366.86 	& 12  	 & b     	& 7.40  	& 4366.90 & 37  	& a      & 7.72  \\
4369.27& O\,{\sc ii} 		&$\cdots$  	&$\cdots$&$\cdots$     	&$\cdots$	& 4369.30 & 10  	& c      & 7.80  \\
4395.94& O\,{\sc ii} 		&$\cdots$  	&$\cdots$&$\cdots$     	&$\cdots$	& 4395.95 & 11  	& c      & 7.69  \\
4414.91& O\,{\sc ii} 		& 4414.90 	& 26  	 & a     	& 7.53  	& 4414.91 & 70  	& a      & 7.86  \\
4416.98& O\,{\sc ii} 		& 4416.98 	& 20  	 & a     	& 7.64  	& 4416.98 & 52  	& a      & 7.89  \\
4452.38& O\,{\sc ii} 		&$\cdots$	&$\cdots$&$\cdots$      &$\cdots$       & 4452.37 & 16  	& b      & 7.90  \\
4590.97& O\,{\sc ii} 		& 4590.98 	& 11  	 & b     	& 7.41  	& 4590.99 & 43  	& a      & 7.66  \\
4596.17& O\,{\sc ii} 		& 4596.17 	& 8    	 & c     	& 7.38  	& 4596.19 & 33  	& a      & 7.61  \\
4609.44& O\,{\sc ii}$^a$	&$\cdots$	&$\cdots$&$\cdots$ 	&$\cdots$	& 4609.44 & 12  	& c      & 7.36  \\
4638.86& O\,{\sc ii} 		& 4638.87  	& 14  	 & a     	& 7.54  	& 4638.87 & 50  	& a      & 7.87  \\
4641.81& O\,{\sc ii} 		& 4641.83  	& 27  	 & a     	& 7.52  	& 4641.83 & 79  	& a      & 7.85  \\
4649.14& O\,{\sc ii}$^a$$^b$	& 4649.14 	& 39  	 & a     	& 7.53	      	& 4649.16 & 102 	& a      & 7.86  \\
4650.84& O\,{\sc ii}$^a$$^b$	& 4650.84 	& 14  	 & a     	&$\cdots$ 	& 4650.86 & 48  	& a      &$\cdots$\\
4661.64& O\,{\sc ii} 		& 4661.63 	& 18   	 & a     	& 7.61  	& 4661.65 & 50  	& a      & 7.81  \\
4673.74& O\,{\sc ii} 		&$\cdots$	&$\cdots$&$\cdots$ 	&$\cdots$	& 4673.75 & 13  	& b      & 7.90  \\
4676.24& O\,{\sc ii} 		& 4676.21 	& 16  	 & b     	& 7.62  	& 4676.25 & 44  	& a      & 7.83  \\
4699.00& O\,{\sc ii}$^a$$^b$	&$\cdots$  	&$\cdots$&$\cdots$      &$\cdots$	& 4699.03 & 34  	& c	 & 7.63  \\
4699.22& O\,{\sc ii}$^a$$^b$	&$\cdots$  	&$\cdots$&$\cdots$ 	&$\cdots$	& 4699.24 &$\cdots$ 	& b	 &$\cdots$\\
4705.35& O\,{\sc ii} 		&$\cdots$  	&$\cdots$&$\cdots$     	&$\cdots$	& 4705.37 & 37  	& a	 & 7.70  \\
4710.01& O\,{\sc ii}$^a$ 	&$\cdots$  	&$\cdots$&$\cdots$     	&$\cdots$	& 4710.04 & 11  	& c	 & 7.93  \\
\hline

\end{tabular}
\end{minipage}
\end{table*}

\begin{table*}
\begin{minipage}{\textwidth}
\addtocounter{table}{-1}
\caption{(continued)}
\begin{tabular}{@{}llccccccccc}
\hline
\multicolumn{2}{c}{Identification}&\multicolumn{4}{c}{Barnard\,29}&\multicolumn{4}{c}{ROA\,5701}\\
Rest $\lambda$ & Species 	&Observed $\lambda$ & W & Quality$^\dagger$ & Abundance$^\diamond$ &Observed $\lambda$ & W & Quality$^\dagger$ & Abundance$^\diamond$\\ 
(\AA)		&		& (\AA)			& (m\AA)	& & & (\AA)	& (m\AA)	\\
\hline\\

4890.93& O\,{\sc ii} 		&$\cdots$  	&$\cdots$&$\cdots$	&$\cdots$	& 4890.87 & 11  	& a	 & 7.79  \\
4906.83& O\,{\sc ii} 		&$\cdots$  	&$\cdots$&$\cdots$	&$\cdots$	& 4906.84 & 15  	& c	 & 7.60  \\
4924.53& O\,{\sc ii} 		&$\cdots$  	&$\cdots$&$\cdots$	&$\cdots$	& 4924.53 & 28  	& a	 & 7.91  \\
4941.07& O\,{\sc ii} 		&$\cdots$  	&$\cdots$&$\cdots$	&$\cdots$	& 4941.10 & 12  	& c	 & 7.90  \\
4943.00& O\,{\sc ii}$^a$	&$\cdots$	&$\cdots$&$\cdots$	&$\cdots$	& 4943.01 & 21  	& a      & 7.94  \\

4481.13& Mg\,{\sc ii}$^a$$^b$	& 4481.12 	& 19  	 & a     	& 6.10  	& 4481.16 & 10  	& b      & 6.15  \\
4481.33& Mg\,{\sc ii}$^a$$^b$	& 4481.32 	& 11  	 & b     	&$\cdots$       & 4481.34 & 7   	& b      &$\cdots$\\

4128.05& Si\,{\sc ii}		& 4128.04 	& 11   	& c    		& 6.45 		&$\cdots$ &$\cdots$	&$\cdots$&$\cdots$\\
4552.62&Si\,{\sc iii}		& 4552.60 	& 71   	& a    		& 6.45		& 4552.63 & 67  	& a	 & 6.18  \\
4567.82&Si\,{\sc iii}		& 4567.86 	& 53   	& a    		& 6.43		& 4567.86 & 50  	& a	 & 6.18  \\
4574.76&Si\,{\sc iii}		& 4574.77 	& 30   	& a    		& 6.48		& 4574.77 & 23  	& a	 & 6.16  \\
5739.74&Si\,{\sc iii}		& 5739.76 	& 26   	& b    		& 6.59		& 5739.78 & 23  	& b 	 & 5.96  \\
4088.85& Si\,{\sc iv}		&$\cdots$  	&$\cdots$&$\cdots$	&$\cdots$	& 4088.89 & 33  	& a	 & 6.15  \\
4116.10& Si\,{\sc iv}		&$\cdots$  	&$\cdots$&$\cdots$	&$\cdots$	& 4116.12 & 21  	& b	 & 6.17  \\ 

4512.54&Al\,{\sc iii}		& 4512.58 	& 7    	& b    		& 5.13 		&$\cdots$ &$\cdots$	&$\cdots$&$\cdots$\\
4529.18&Al\,{\sc iii}		& 4529.18 	& 11   	& b    		& 5.15		&$\cdots$ & $<$5	&$\cdots$& $<$4.87\\ 
	
4253.59& S\,{\sc iii}		& 4253.57 	& 7    	& a    		& 5.72		& 4253.60 & 14  	& b	 & 5.62  \\

4419.60& Fe\,{\sc iii}		& 4419.60	& 3	& c		& 6.03		&$\cdots$ & $\cdots$	&$\cdots$&$\cdots$\\
5156.12& Fe\,{\sc iii}		& 5156.09	& 5	& c		& 6.08		&$\cdots$ & $\cdots$	&$\cdots$&$\cdots$\\
5833.93& Fe\,{\sc iii}		& 5833.94	& 4	& c		& 6.21		& 5833.99 & $<$6	& c	 & $<$6.10  \\

\hline

\end{tabular}
\medskip\\
$^\dagger$ a indicates that the equivalent width measurement should be accurate to better than 
$\pm$10 per cent, b to better than $\pm$20 per cent, and c to less than $\pm$20 per cent.
\\
$^\diamond$ Logarithmic abundance [M/H] on the scale log[H] = 12.00.\\
$^a$ Treated as blends in {\sc tlusty}.\\
$^b$ Equivalent widths of adjacent lines were combined to produce the corresponding abundances shown.
\end{minipage}
\end{table*}

Table \ref{tab_ab} contains a summary of the abundances for each species, along with the associated error estimates and the number of lines observed. 
Also included are the abundances previously derived for the objects \citep{{con94},{dix98},{moe98},{tho06}}, and those of normal B-type stars, both in the Galactic cluster NGC\,6611 \citep{hun06} and in the local field \citep{{kil92},{kil94}}. The former were determined using the same {\sc tlusty} code as employed in the present paper. Also included are typical interstellar medium (ISM) abundances for the elements concerned, as listed in \citet*{wil00}. The C, N and O abundances are taken from \citet{car96} and \citet*{{mey97},{mey98}} respectively, with all other elements given by \citet{sno96}. The Al abundance, however, is taken as 70 per cent of the adopted solar abundance of  \citet{wil00}, as ISM abundances appear to be reduced by 20 -- 30 per cent with respect to solar abundances \citep{{sno96},{sav96}}.

The errors associated with the absolute abundances (see Table \ref{tab_ab}) arise from systematic and random errors (full details can be found in \citealt{hun05}).  Random errors are connected with the data analysis, for example: line fitting errors, oscillator strengths and observational uncertainties. 
The random uncertainty is taken as the standard deviation in the abundances of a given species, divided by the square root of the number of lines observed for 
that species. Where only one line was detected, the standard deviation is that of the best observed species, e.g. O\,{\sc ii}. 
Systematic errors arise from the uncertainties in the atmospheric parameters, and are estimated by varying each parameter by its associated errors. 
The total uncertainty is the square root of the sum of the squares of the random and systematic errors. 
We note that the scatter of individual abundances for each species is reasonably small, with any discrepancies probably arising from poorly determined lines and possible errors in atomic data.

\begin{table*}
\begin{minipage}{\textwidth}
\begin{center}
\caption{Absolute abundances for Barnard\,29 and ROA\,5701 from their optical spectra, along
with those previously derived by \citet{{con94},{dix98},{moe98}} and \citet{tho06}. Also listed
are abundances for normal B-type stars from the local field \citep{{kil92},{kil94}} and the Galactic cluster NGC\,6611 \citep{hun06}, plus values for the interstellar medium \citep{{car96},{mey97},{mey98},{sno96},{wil00}}. Bracketed values are the number of lines of each species observed.}
\label{tab_ab}
\begin{tabular}{@{}l cccccccccccc}
\hline
Species 	&\multicolumn{5}{c}{Barnard\,29}&\multicolumn{4}{c}{ROA\,5701} &\multicolumn{2}{c}{B-type} & ISM\\
		&\multicolumn{2}{c}{This paper}&Con\,94	&Dix\,98 &Moe\,98$^\star$ &\multicolumn{2}{c}{This paper} &Moe\,98$^\star$&Tho\,06& Kil\,92/94 & Hun\,06&\\
\hline \\
C\,{\sc ii} 	& $<$6.64	 &(1)   & $<$6.70	& 6.15	      	&$\cdots$	& $<$6.54 	 &(1) & $<$5.85	& $<$6.51& 8.20      & 7.95    	& 8.38	 \\
N\,{\sc ii} 	&7.43 $\pm$ 0.12 &(17)  &    7.30	&$\cdots$	&$\cdots$	&7.06 $\pm$ 0.12 &(9) &  6.86 	&    7.05& 7.69      & 7.59	& 7.88   \\
O\,{\sc ii} 	&7.50 $\pm$ 0.24 &(17)  &    7.60	&$\cdots$     	&$\cdots$      	&7.76 $\pm$ 0.10 &(41)&    7.97 &    7.75& 8.55      & 8.55 	& 8.69   \\
Mg\,{\sc ii}	&6.10 $\pm$ 0.15 &(1)	&    6.00	&$\cdots$     	&$\cdots$      	&6.15 $\pm$ 0.21 &(1) & $\cdots$&    6.27& 7.38      & 7.32 	& 7.40   \\
Si\,{\sc ii} 	&6.45 $\pm$ 0.20 &(1)	&$\cdots$	&$\cdots$     	&$\cdots$      	&$\cdots$  	 &    & $\cdots$&$\cdots$& $\cdots$  &$\cdots$ 	&$\cdots$\\
Si\,{\sc iii} 	&6.45 $\pm$ 0.26 &(3)	&    6.30	&$\cdots$     	&$\cdots$      	&6.17 $\pm$ 0.20 &(3) &    6.14 &    5.94& 7.28      & 7.41 	& 7.27   \\
Si\,{\sc iv} 	& $\cdots$ 	 &      &$\cdots$	&$\cdots$	&$\cdots$	&6.16 $\pm$ 0.40 &(2) & $\cdots$&    5.95& $\cdots$  &$\cdots$ 	&$\cdots$\\
Al\,{\sc iii} 	&5.14 $\pm$ 0.10 &(2)	&    5.19	&$\cdots$	&$\cdots$	& $<$4.87	 & (1)& $\cdots$& $<$5.21 & 6.22     &$\cdots$ 	& 6.33   \\
S\,{\sc iii} 	&5.72 $\pm$ 0.26 &(1)	& $<$6.29   	& 5.34		&$\cdots$	&5.62 $\pm$ 0.23 &(1) & $\cdots$&    5.71& 7.25	     &$\cdots$ 	& 7.09   \\
Fe\,{\sc iii} 	&6.07 $\pm$ 0.12 &(3)   & $<$6.70	& 5.28	      	& 5.30 		& $<$6.10 	 & (1)& 4.79	&$\cdots$& $\cdots$  & 7.50    	& 7.43   \\
\hline
\end{tabular}
\end{center}
\medskip
$^\star$ average of curve-of-growth and spectrum synthesis values.
\\
\end{minipage}
\end{table*}

A normal He abundance has been assumed throughout the analysis of both stars, which was tested by over-plotting theoretical spectra on the observations in regions containing He\,{\sc i} absorption features. There is good agreement to within the uncertainties of the atmospheric parameters, suggesting this is a valid assumption. When lines are badly blended with hydrogen or helium lines, they are not included in the non-LTE calculations.

For both stars, the best observed species are N\,{\sc ii} and O\,{\sc ii}, with the majority of equivalent widths believed to be accurate to better than 20 per cent. If the lines of poorer quality are removed, the abundance estimates remain unchanged (within the error bars), indicating 
that data of poorer quality do not significantly affect the abundances obtained. 

The C abundances for both Barnard\,29 and ROA\,5701 in Table \ref{tab_ab} are quoted as upper limits. Due to the quality of the spectra, it 
is was only possible to place upper limits on the equivalent widths of the C\,{\sc ii} lines at 4267 and 6578 \AA. 
\citet{sig96} finds that the 4267 \AA\ multiplet gives a limiting precision of $\sim$\,$\pm$ 0.2 dex, over a $T_{\rm eff}$ range of 15000 -- 31000 K (as reproduced by \citealt{nie06}). However, for $T_{\rm eff}$ $<$ 25000 K, the line at 6578 \AA\ is somewhat more reliable than that at 4267 \AA. Therefore, the upper limits given in Table \ref{tab_ab} are based on the 6578 \AA\ line alone.

The Mg\,{\sc ii} doublet at 4481 \AA\ was observed and treated as a blend within {\sc tlusty}, and for both stars the derived abundances appear reliable. We note that the Si abundance estimates are sensitive to the adopted atmospheric parameters. For example, for 
Barnard\,29 a change in $T_{\rm eff}$ of $\pm$1000 K leads to a variation of $\pm$0.15 dex in the 
abundance of Si\,{\sc ii} and $\pm$0.21 for Si\,{\sc iii}. In the case of ROA\,5701, the same temperature change produces an abundance variation of
$\pm$0.10 dex for Si\,{\sc iii} and $\pm$0.34 dex for Si\,{\sc iv}.

The Si\,{\sc iii} line at 5739 \AA\ was observed in both stars. However, as there are no non-LTE grids available for this spectral region, theoretical LTE equivalent widths were generated, as for S\,{\sc iii}, Al\,{\sc iii} and Fe\,{\sc iii}. The resulting LTE abundances are 6.59 dex for Barnard\,29 and 5.96 dex for ROA\,5701, which compare well to the Si\,{\sc iii} results in Table \ref{tab_ab}. 
\citet{gie92} and \citet{kil94} found that LTE abundances agree to 0.2 dex with corresponding non-LTE abundances for B-type stars, consistent with the results here. However, as the 5739 \AA\ abundances are in LTE, they are not included in the relevant non-LTE Si\,{\sc iii} results listed in Table \ref{tab_ab}. 

The S\,{\sc iii} 4254 \AA\ line was used to derive the S abundance in both Barnard\,29 and ROA\,5701. 
As previously noted, these abundances are based on theoretical LTE S\,{\sc iii} equivalent widths. \citet{hun05} find that non-LTE effects appear to be negligible for this species. 
Additionaly, we have carried out test calculations and find little difference between the S\,{\sc iii} spectrum predicted using LTE or non-LTE. Hence our use of an LTE approach is probably valid. 

Two Al\,{\sc iii} lines, at 4512 and 4259 \AA, were observed for Barnard\,29. No Al features were detected for ROA\,5701, but an upper limit on the Al abundance was set using the 4529 \AA\ line. 
The Al abundances are treated in LTE, but should be reliable as \citet{hun05} state that when using the dominant ion stage, LTE abundance calculation will give good agreement with non-LTE abundances.   

\begin{figure*}
\begin{minipage}{\textwidth}
\includegraphics[angle=270,width=\textwidth]{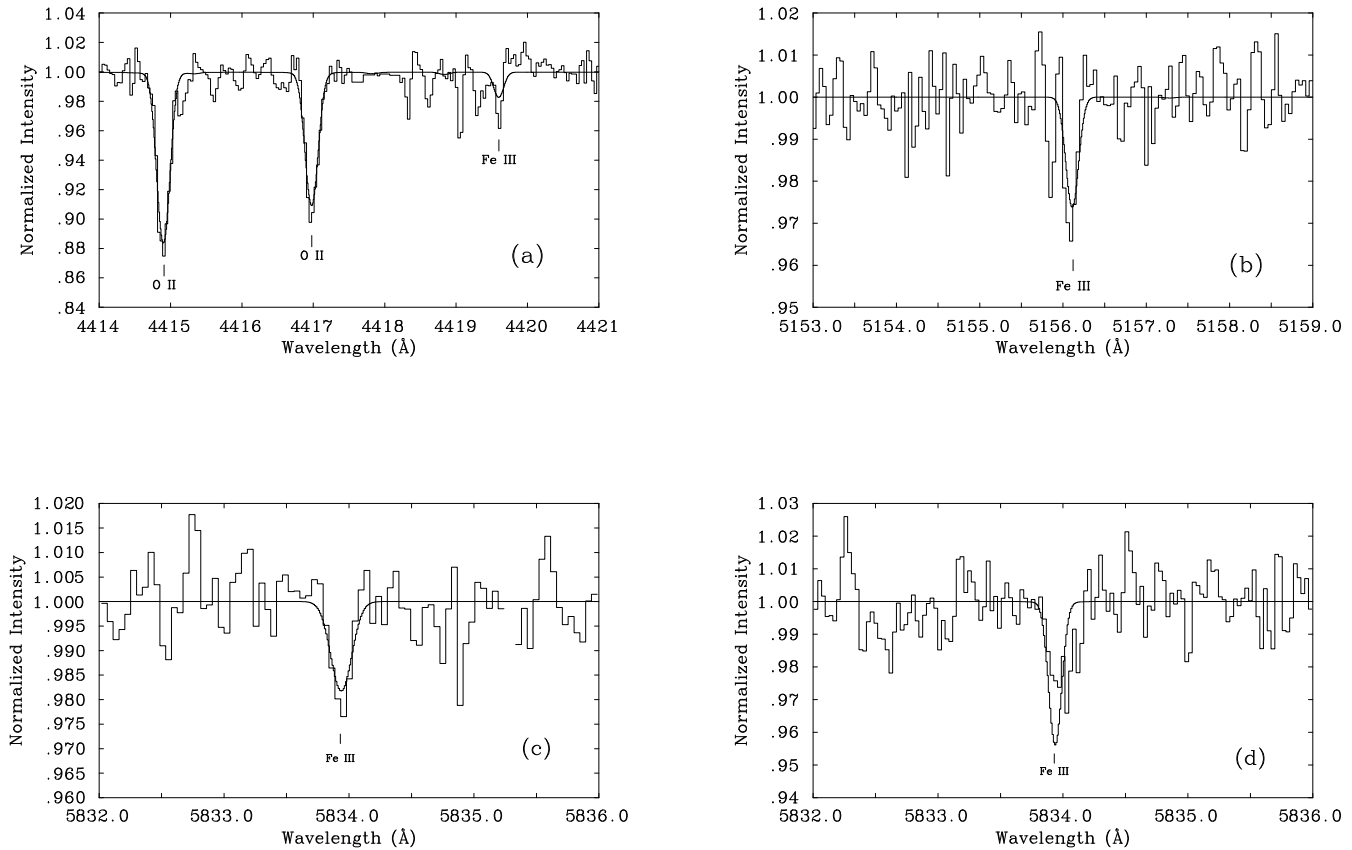}
\caption{Observed and theoretical line profiles for Fe\,{\sc iii} in Barnard\,29 (a,b and c) and ROA\,5701 (d). The theoretical profiles (smooth lines) are the best fits to the observed features, and have been calculated at abundances of (a) O\,{\sc ii} 4414 \AA\ = 7.53 dex, O\,{\sc ii} 4416 \AA\ = 7.64 dex, and Fe\,{\sc iii} 4419 \AA\ = 6.03 dex; (b) Fe\,{\sc iii} 5156 \AA\ = 6.08 dex; (c) Fe\,{\sc iii} 5833 \AA\ = 6.21 dex, and (d) Fe\,{\sc iii} 5833 \AA\ = 6.10 dex.}
\label{fig_fe_op}
\end{minipage}
\end{figure*}

Three Fe\,{\sc iii} lines were observed at 4419, 5156 and 5833 \AA\ in Barnard\,29, with  one in ROA\,5701 at 5833 \AA. The S/N of the local continuum for the lines are $\sim$130, 155, 150 and 110 per pixel, respectively. LTE calculations were used to determine the Fe abundance, and Fig. \ref{fig_fe_op} shows theoretical profiles corresponding to the measured abundances, over-plotted onto the observed spectra of Barnard\,29 (a,b and c) and ROA\,5701 (d). 
As can be seen from the figure, some of the features are difficult to distinguish from the noise. Indeed, Fig. \ref{fig_fe_op}(a) appears to show the 4419 \AA\ line in Barnard\,29, but numerous noise features of similar size to the 4419 \AA\ line can also be seen, which may be due to observational effects or the reduction process. Unless these additional features are real (which is unlikely), then the Fe\,{\sc iii} line may also not be real and hence the abundance derived should be used with care. In addition, the 4419 \AA\ line appears to be narrower than other weak stellar absorption features in the spectrum, as may be seen from the profile fit in Fig. \ref{fig_fe_op}(a), where the theoretical feature is wider than that observed. However, 4419 \AA\ gives a similar abundance to the apparently more reliable 5156 \AA\ line, and so all three observed Fe\,{\sc iii} transitions are used in determining the iron abundance for Barnard\,29. 
Also, these lines have the largest gf-values in their respective multiplets, and are among the strongest unblended Fe\,{\sc iii} lines in the optical spectral region. 
For ROA\,5701, the sole Fe\,{\sc iii} line observed at 5833 \AA\ does not appear reliable (see Fig. \ref{fig_fe_op}(d)) and hence the abundance is taken as an upper limit.

As the observed optical iron lines are of low quality, ultraviolet spectra were also employed to determine Fe abundances for both Barnard\,29 and ROA\,5701. This is discussed below.

\subsection{Iron abundances from ultraviolet spectra}
\label{sec_uv2}

To determine an iron abundance from the GHRS/HST spectrum, a goodness-of-fit method was used (\citealt{{rya02},{duf06}}, Dufton (in preparation)). Theoretical spectra were generated for the regions observed by the GHRS/HST, using non-LTE {\sc tlusty} model atmospheres and appropriate stellar atmospheric parameters. The model grid for low metallicity regions 
was again used, and the underabundance of elements (heavier than He, up to and including Zn), was varied in each spectrum (in steps of 0.1 dex from 0.0 to --2.5 dex) relative to the solar estimates adopted in {\sc tlusty} \citep{gre98}. 
Light elements (C, N, O, Mg, Si) were treated in non-LTE while all other elements were in LTE. Dufton (in preparation) found that $\sim$80 per cent of absorption in the ultraviolet spectral region covered by our GHRS/HST data are due to Fe,
with more than 95 per cent from Fe-group elements, therefore the calculations are taken as being in LTE.
Iron dominates the absorption, therefore any variation in the relative element abundances should not significantly affect the theoretical spectra. 

\begin{figure*}
\begin{minipage}{\textwidth}
\includegraphics[angle=270,width=\textwidth]{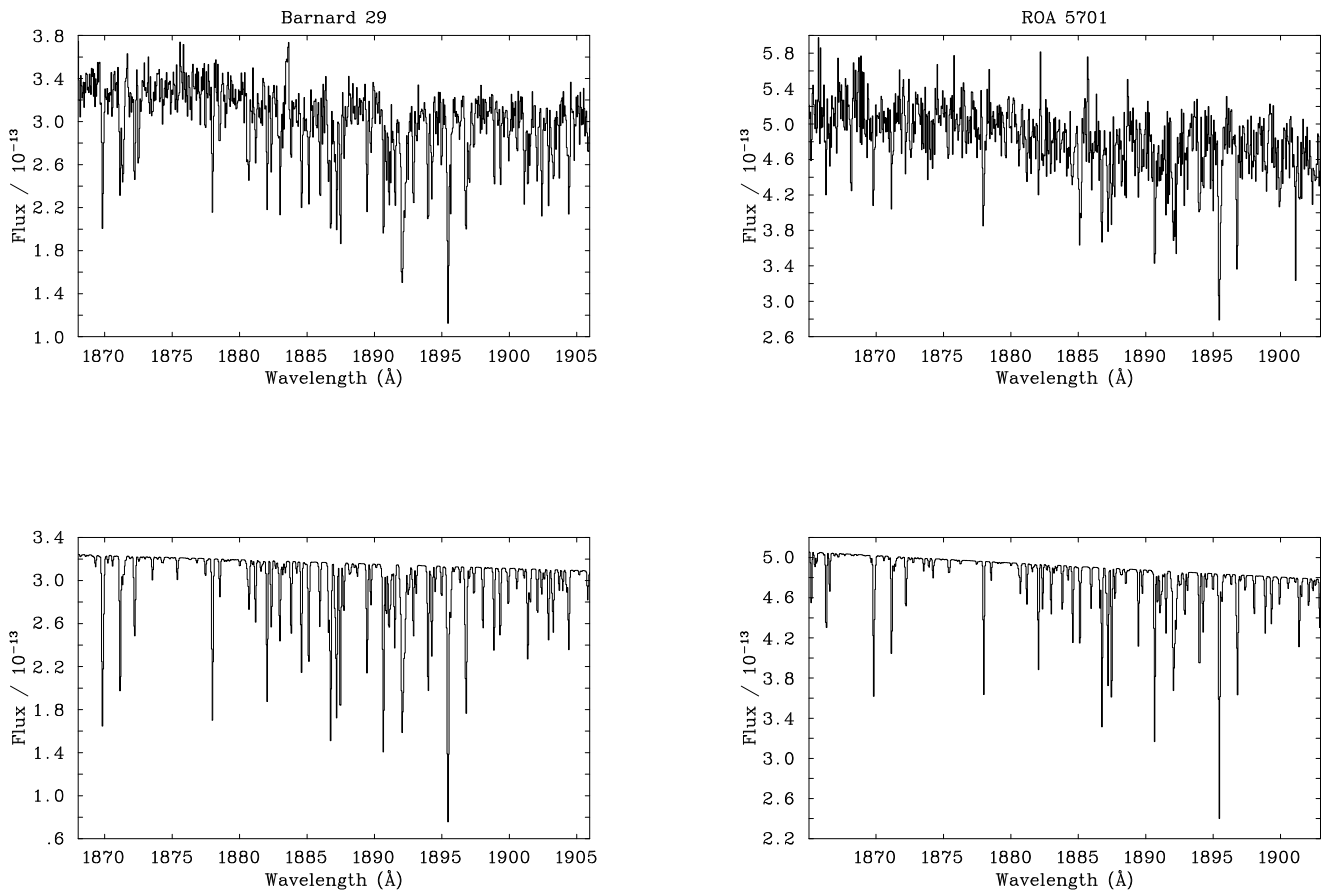}
\caption{The upper panels show the observational data for Barnard\,29 and ROA\,5701 obtained
with GHRS/HST, while the lower panels contain
the theoretical spectra for [Fe/H] = --2.1 and --2.4, respectively, relative to the Galactic value of [Fe/H] = 7.5.}
\label{fig_all}
\end{minipage}
\end{figure*}

Due to the numerous absorption features in the ultraviolet spectral region, continuum placement was difficult,
as may be seen from the GHRS/HST observations of Barnard\,29 and ROA\,5701 in Fig. \ref{fig_all}. Therefore, the continuum was not normalized, instead the flux calibrated spectrum was examined. 
This was achieved by smoothing the theoretical spectra with a Gaussian function (to allow for instrumental broadening), followed by binning to the exact wavelength range and interval as the observed spectra, and then scaling to the same mean flux as that observed. 

A goodness-of-fit was calculated by finding the sum of the squares of the difference between the observed and theoretical spectra for a range of metallicities. A plot of this produced a parabolic shape, with the model with the best goodness-of-fit situated at the 
minimum of the parabola.  
This method removes the need to normalize the observed spectrum, thus reducing errors associated with fitting the continuum.
To test if different spectral intervals would affect the abundance estimate, smaller wavelength ranges were selected and analysed. No significant change was found in the goodness-of-fit and associated metallicity value.

Results of the minimization procedure are shown in Table \ref{tab_uv}, along with iron abundances derived by \citet{moe98} and \citet{dix98}. Errors were calculated using a similar method as for the optical spectra. Systematic errors were obtained by varying the atmospheric parameters by their respective errors, while random errors were estimated by over-plotting theoretical spectra around the minimum and observing where the spectrum no longer matched the strongest observed features. The total uncertainty was then calculated and is shown in Table \ref{tab_uv}.

The theoretical spectra that produced a minimum are shown, together with the observed spectra, in Fig. \ref{fig_all}. 
For Barnard\,29, the goodness-of-fit indicates an [Fe/H] = --2.1. In general, the theoretical spectrum matched many of the individual absorption features in the 
observational data, supporting the abundance extimates for this star. ROA\,5701 was less convincing in the fitting of the observed spectrum, with a minimum produced for [Fe/H] = --2.4 dex. In particular, the spectrum has fewer identifiable absorption features compared to Barnard\,29, making the estimation of the metallicity unreliable. 

In Fig. \ref{fig_all}, there is an emission feature at $\sim$1883 \AA\ in the spectrum of Barnard\,29, which appears real and resembles part of a P Cygni profile. However, the apparent line does not have a strong associated absorption feature and is therefore thought to be an artefact of the reduction process. If it is removed from the goodness-of-fit analysis, there is no change in the abundance obtained. The feature was not observed in ROA\,5701.

\begin{table*}
\begin{minipage}{\textwidth}
\begin{center}
\caption{Values of [Fe/H] derived for Barnard\,29 and ROA\,5701 relative to the Galactic abundance, 
along with previous estimates.}
\label{tab_uv}
\begin{tabular}{@{}lcccccccccc}
\hline
Star		& This paper    	& \citet{moe98} &\citet{dix98}	\\
\hline\\
Barnard\,29	&   --2.1 $\pm$ 0.1	& --2.2    & --2.2	\\
ROA\,5701	&   --2.4 $\pm$ 0.3	& --2.7    & $\cdots$ 	\\
\hline
\end{tabular}
\end{center}
\end{minipage}
\end{table*}

\section{Discussion}

Barnard\,29 and ROA\,5701 have been the subject of numerous studies, several of which have included
determinations of atmospheric parameters and chemical compositions. We discuss these below in the
context of the present results.

\subsection{Comparison of atmospheric parameters}

\subsubsection{Barnard\,29}

Previous studies of Barnard\,29 have produced similar atmospheric parameters to those found in this paper. For example, \citet{aue74} used non-LTE models with optical spectra to derive $T_{\rm eff}$ = 22500 $\pm$ 1200 K and log~{\em g} = 3.0 $\pm$ 0.15 dex, while 
\citet{deb85} employed ultraviolet spectra from the {\it IUE} satellite and ultraviolet colours, 
finding $T_{\rm eff}$ = 19000 $\pm$ 1000 K and log~{\em g} = 3.2 dex. These spectra were again employed by \citet{ade94}, along with optical spectra of H$\gamma$, to determine LTE atmospheric parameters of $T_{\rm eff}$ = 20250 K and log~{\em g} = 3.15 dex. \citet{con94} used LTE models to analyse {\it IUE} low-resolution data and WHT/UES optical data, resulting in $T_{\rm eff}$ = 20000 $\pm$ 1000 K, log~{\em g} = 3.0 $\pm$ 0.1 dex and $\xi$ = 10 $\pm$ 2 km s$^{-1}$. 
\citet{dix98} took a sample of the {\it IUE} spectra used by \citet{con94}, along with {\it ORFEUS-SPAS\,II} spectra, and with LTE techniques obtained the same surface gravity as \citet{con94}, but an increased temperature estimate of $T_{\rm eff}$ = 21000 $\pm$ 1000 K. 

The previously derived effective temperatures and surface gravities are in good agreement with those found in this paper, within the errors, with the exception of the \citet{aue74} temperature, which may be due to using different non-LTE models.
\citet{con94} derived a larger microturbulence than in the present paper, presumably due to 
using LTE models rather than our non-LTE approach. 
Also \citet{con94} only employed O\,{\sc ii} lines in their determination of $\xi$, compared to our use of both the Si\,{\sc iii} multiplet and a variety of different 
multiplets of O\,{\sc ii}. Given the high quality of our observational data combined with the complexity of our non-LTE models, we 
believe that our atmospheric parameters for Barnard\,29 are probably the most reliable estimates to date.

\subsubsection{ROA\,5701}

ROA\,5701 was studied recently by \citet{tho06} using the same non-LTE models and methods as in 
the present paper, resulting in atmospheric parameters of $T_{\rm eff}$ = 25000 $\pm$ 1000 K, 
log~{\em g} = 3.2 $\pm$ 0.2 dex and $\xi$ = 5 $\pm$ 5 km s$^{-1}$. 
\citet{moe98} employed ESO CASPEC optical spectra, along with the GHRS/HST data used here, and found
$T_{\rm eff}$ = 22000--24500 K, log~{\em g} = 3.2--3.4 dex. In addition, they derived
$\xi$ = 2--3 or 20 km s$^{-1}$, the former using Fe\,{\sc iii} transitions and the latter those of
O\,{\sc ii}. Their use of LTE models, compared to the non-LTE approach in the present paper, 
may be the reason for the differences in the derived microturbulence values. 
The present microturbulence estimate is smaller than that of \citet{tho06}, 
probably due to improved equivalent width measurements and surface gravity estimate, 
although it is still within the uncertainty quoted. 

Other studies of ROA\,5701 include \citet{nor74}, who analysed narrow-band and UBV photometry to obtain 
$T_{\rm eff}$ = 26600 K and log~{\em g} = 3.5 dex, and \citet{cac84} who used {\it IUE} spectra to derive $T_{\rm eff}$ = 24000 K. 
These studies are in agreement with the parameters derived here, with the exception of the $T_{\rm eff}$ estimate of \citet{nor74}. \citet{cac84} suggested that the different values of (B-V) used by \citet{nor74} are responsible for their larger temperature estimate. The results derived here are hence generally consistent with previous values, but once again are believed to be more reliable due to the improved observational data and non-LTE models employed in our analysis.

\subsection{Comparison of chemical compositions}

Table \ref{tab_ab} shows the absolute abundances derived for Barnard\,29 and ROA\,5701, along with those previously determined by \citet{{con94},{moe98}} and \citet{dix98} for Barnard\,29, and \citet{moe98} and \citet{tho06} for ROA\,5701.
However, absolute abundances may contain systematic errors, due for example to inaccuracies 
in the atmospheric parameters and atomic data, in particular the f-values. 
In order to minimize these errors, a detailed line-by-line differential abundance analysis can be performed, with respect to a young B-type star of known chemical composition (Population\,{\sc i}) which has similar atmospheric parameters. Therefore, in Table \ref{tab_diff} the 
abundances of Barnard\,29 and ROA\,5701 are compared to those of young B-type stars in the Galactic cluster NGC\,6611 \citep{hun06} and in the local field \citep{{kil92},{kil94}}. For the former, results have been determined using the same non-LTE model atmosphere calculations as in the present paper. The latter uses different model atmosphere calculations and hence differential abundances using this source may not be as reliable. 
The standard error in the differential abundance estimates for Barnard\,29 and ROA\,5701, as compared with young B-type stars, are typically $\pm$0.2 dex. These errors do not include any systematic errors due to the atmospheric parameters as these are assumed to cancel due to the use of the same methods for both samples. 
Note that where we did not have access to results for individual absorption lines, differential abundance values are based on the mean abundance estimates listed, with a minimum error of $\pm$0.2 dex. 

It is not possible to perform a differential analysis for Fe\,{\sc iii} ultraviolet lines using young B-type stars, as these
are affected by line crowding and are often saturated. 
The strengths of the Fe\,{\sc iii} absorption lines are only weakly dependent on the effective temperature and surface gravity,
so they should not be a significant source of error. 
In order to test the dependence on microturbulence for young B-type stars, the 
Galactic star HR\,1886 was examined using STIS spectra taken from the HST archive. 
This star was chosen as its $T_{\rm eff}$ = 24000 K and log~{\em g} = 4.1 dex \citep{ham96} are similar to those of ROA\,5701.  
For a microturbulence of 0 km s$^{-1}$, synthetic spectra were generated and scaled to the same mean flux as the observed spectrum, 
using the methods described in Section \ref{sec_uv2}. The microturbulence was then increased to 
10 km s$^{-1}$, and the process repeated. The change in microturbulence resulted in a change of metallicity of $\sim$1.4 dex, implying that Fe\,{\sc iii} ultraviolet lines in Galactic B-type stars are indeed very sensitive to changes in microturbulence, and are not appropriate for a differential abundance analysis.

\citet{ken94a} found a similar dependence on microturbulence for the standard star HD\,149438 ($\tau$\,Scorpii), 
with a change of $\xi$ from 0 to 10 km s$^{-1}$ yielding a metallicity change of $\sim$0.8 dex.
For Barnard\,29 and ROA\,5701, a $\xi$ change from 0 to 10 km s$^{-1}$ resulted in a metallicity change of $\sim$0.4 and 0.2 dex, respectively. 

\begin{table*}
\begin{minipage}{\textwidth}
\begin{center}
\caption{Differential abundances for Barnard\,29 and ROA\,5701, relative to those in 
young B-type stars from \citet{hun06} where available, supplemented by \citet{{kil92},{kil94}}$^\dagger$. 
Also listed are differential abundances relative to the ISM values \citep{{car96},{mey97},{mey98},{sno96},{wil00}}, and the metallicities of M\,13$^\diamond$ \citep{yon06} and $\omega$\,Cen$^\star$ \citep{ori03}.}
\label{tab_diff}
\begin{tabular}{@{}l cccccccccccccccc} 
\hline
Species 	& \multicolumn{7}{c}{Young B-type stars}					&\multicolumn{2}{c}{ISM}\\
		&\multicolumn{4}{c}{Barnard\,29}	      	&\multicolumn{3}{c}{ROA\,5701} 	&Barnard\,29& ROA\,5701\\
		&This paper&Con\,94 &Dix\,98	 &Moe\,98	&This paper&Moe\,98	&Tho\,06   & \multicolumn{2}{c}{This paper} \\
\hline \\
C  		&$<$--1.3 	&$<$--1.3 	& --1.8      	&$\cdots$	& $<$--1.4 	& $<$--2.1  	& $<$--1.4 & $<$--1.7   & $<$--1.8\\
N  		&   --0.3 	&   --0.3 	&$\cdots$	&$\cdots$      	&    --0.6 	&    --0.7   	&    --0.6 &    --0.5   &    --0.8\\
O  		&   --1.1 	&   --1.0 	&$\cdots$	&$\cdots$	&    --0.8 	&    --0.6   	&    --0.8 &    --1.2   &    --0.9\\
Mg  		&   --1.2 	&   --1.3 	&$\cdots$	&$\cdots$      	&    --1.2 	& $\cdots$   	&    --1.1 &    --1.3   &    --1.3\\
Si  		&   --1.0 	&   --1.1 	&$\cdots$	&$\cdots$	&    --1.2 	&    --1.3   	&    --1.5 &    --0.8   &    --1.1\\
Al$^\dagger$  	&   --1.1 	&   --1.0 	&$\cdots$	&$\cdots$      	& $<$--1.4 	& $\cdots$   	& $<$--1.0 &    --1.2   & $<$--1.5\\
S $^\dagger$ 	&   --1.5 	&$<$--1.0 	& --1.9	 	&$\cdots$     	&    --1.6 	& $\cdots$	&    --1.5 &    --1.4   &    --1.5\\
Fe (Optical)	&   --1.4 	&$<$--0.8 	&$\cdots$ 	&$\cdots$	& $<$--1.4 	& $\cdots$ 	&$\cdots$  &    --1.4   & $<$--1.4\\
Fe (UV)		&   --2.1 	&$\cdots$ 	& --2.2     	& --2.2		&    --2.4 	&    --2.7  	&$\cdots$  &    --2.1   &    --2.1\\
Cluster [Fe/H]	& \multicolumn{4}{c}{--1.6$^\diamond$}		& \multicolumn{3}{c}{--1.6$^\star$}			  \\ 
\hline
\end{tabular}
\end{center}
\medskip
Note: 
Con\,94 = \citet{con94},
Dix\,98 = \citet{dix98},
Moe\,98 = \citet{moe98},
Tho\,06 = \citet{tho06}.
\end{minipage}
\end{table*}

\subsubsection{Barnard\,29}
\label{sec_b29chem}

Differential abundances for Barnard\,29, relative to young B-type stars, 
are given in Table \ref{tab_diff}, along with the results of \citet{{con94}, {dix98}} and \citet{moe98}. 
As stated previously, \citet{con94} used {\it IUE} and WHT/UES data with LTE models in their
analysis. Their upper limit for the C abundance agrees with that found in this paper, while N, O, Mg, Si and Al are $\sim$0.1 dex lower. 
Reasons for the differences in abundance could include different atmospheric parameters, 
improved equivalent width measurements (as a consequence of our higher S/N observations), and our use of non-LTE model atmosphere codes. 
To investigate if our non-LTE treatment may be responsible for the differences, we have calculated LTE abundances for the elements, with the exception of Al which is already in LTE.
This results in abundances of $<$5.93 for C, 7.40 for N, 7.75 for O, 5.93 for Mg, 6.78 dex for Si. The N, O and Si abundances are still larger than the \citet{con94} abundances, while C and Mg are lower.

The lower microturbulence used in the present analysis may be a further source of the differences. Adopting a microturbulence of 10 km s$^{-1}$ from \citet{con94}, LTE abundances of 7.30 for N, 7.62 for O and 6.47 dex for Si are obtained,
while the other elements abundances do not significantly change. The values are now in good agreement with \citet{con94}, with the exception of C, suggesting that the microturbulence is a factor. The discrepancy for C may be due to the use of different lines. 
However, the results in the present paper should be more reliable due to our use of non-LTE techniques and improved observational data.

\citet{dix98} used a far-ultraviolet spectrum from {\it ORFEUS-SPAS\,II}, fitting LTE synthetic models 
to the data, similar to the method used here. They determined abundances for the elements He, N, O, Al and S, finding results consistent with \citet{con94} but with larger error bars, suggesting the use of lower resolution data as the reason for this. 
The C abundance of \citet{dix98} was found using two C\,{\sc iii} lines, and is --0.5 dex lower than the upper limit found here. 
Their S abundance was derived from lines of S\,{\sc iii} and S\,{\sc iv}, with some scatter 
due to blending with nearby features. These authors also observed S\,{\sc ii} lines, but these were not
included in the analysis as they appeared to be affected by non-LTE. The S abundance of Dixon \& Hurwitz is $\sim$0.4 dex lower than our value, 
which may be due to the use of different spectral wavelength regions. 

An iron abundance of [Fe/H] = --2.2 was determined by both \citet{dix98} and \citet{moe98}. 
\citet{dix98} found it difficult to determine an Fe abundance, as the spectrum was contaminated by interstellar lines and many of the expected Fe transitions
were in regions of poorly-fitted continuum. They fitted a band of Fe lines rather than individual features, similar to the method used in this paper for the UV observations, noting that their abundance was more than 0.5 dex below the model atmosphere used to derive the synthetic spectra (i.e. [M/H] = --1.5), with tests showing that the model choice had little affect on the abundance.
\citet{moe98} determined an Fe abundance using the same GHRS/HST spectrum employed here. They adopted
the atmospheric parameters of \citet{con94}, and fitted a global continuum to the normalized data, noting
that an uncertainty of 5 per cent in the continuum definition of their spectrum would lead to 
an error of 0.2 dex in the derived abundance.  

In the present paper, an Fe\,{\sc iii} abundance of [Fe/H] = --2.1 is obtained from the UV spectrum and [Fe/H] = --1.4 from the optical. The UV value is well below the metallicity of the cluster ([Fe/H] = --1.6, \citealt{yon06}), 
but is in slightly better agreement compared to the previous UV studies. However, our optical result is consistent with the cluster metallicity, to within the errors. This discrepancy is discussed further in Section \ref{sec_evo}.

\subsubsection{ROA\,5701} 

For ROA\,5701, there is good agreement between the present abundances and those of \citet{tho06}, with C, N and O giving the same differential values. The Si abundance estimates differ by 0.2 dex, and this discrepancy appears to be due to differences in the atmospheric parameters. 
This was tested by calculating non-LTE abundances using the equivalent widths from this paper and the atmospheric parameters from \citet{tho06}. The resulting values are 5.93 dex for Si\,{\sc iii} and 5.88 dex for Si\,{\sc iv}, suggesting the changed atmospheric parameters are a cause for the differences.

The upper limit to the C abundance of \citet{moe98} is lower than that derived in this paper, despite using the same upper limit value for the equivalent width. This difference could be due to large non-LTE effects in C\,{\sc ii}, as proposed by 
\citet{sig96} and hence, the upper limit derived here may be more reliable.
Although the N and O abundances of \citet{moe98} differ by $\sim$0.2 dex, and the Si abundance by $\sim$0.1 dex, from those presented here, again this may be due to non-LTE effects. To test this, 
LTE abundances have been derived, using the {\sc tlusty} grid, resulting in $<$6.23 for C, 7.99 for O, 7.10 dex for N and 6.50 dex for Si. The O abundance is in very good agreement with \citet{moe98}, but the C, N and Si values are still higher. The changes 
could be due to the use of different equivalent width measurements and atmospheric parameters. However, the present results should be more reliable due to our non-LTE analysis.   

No abundances for Mg, Al or S were derived by \citet{moe98}, while \citet{tho06} determined
values for Mg and S, which differ from those found in this paper by $\sim$0.1 dex. This is probably due to the improved S/N observations, which allow better equivalent width measurements; changing the atmospheric parameters had little affect on these abundances.  
The listed Al abundance is still an upper limit, but due to the improved S/N of the spectrum it was possible 
to place a revised lower limit on the equivalent width. 
In general, the abundances quoted in the present paper should be more accurate due to the improved quality of the observational data. 

\citet{tho06} observed two unidentified features at 4512.41 and 4352.22 \AA\ in their data, which were obtained using UCLES at the AAT. However, no features are observed at these wavelengths in the FEROS spectrum, which is of higher resolution and S/N, indicating that the lines in the UCLES data
may not be real. 

\citet{moe98} obtained a differential Fe\,{\sc iii} abundance of [Fe/H] = --2.7 from the GHRS/HST ultraviolet spectrum, compared to [Fe/H] = --2.4 derived here. 
Both values are lower than the metallicity of $\omega$\,Cen, namely [Fe/H] = --1.6 (\citealt{ori03}), although our result is in better agreement with the metallicity of the cluster. 
An optical value [Fe/H] $<$ --1.4 is obtained in the present paper. 
The [Fe/H] abundance of ROA\,5701 is discussed further in Section \ref{sec_evo}. 

\subsection{Comparison with Cluster Giants}

\begin{table}
\begin{center}
\caption{Absolute abundances for Barnard\,29 and ROA\,5701 from their optical spectra. Also tabulated are the abundances of the M\,13 giants (\citealt{coh05}$^a$ and \citealt{yon06}) and those of $\omega$\,Cen, using the main cluster giant population (\citealt{smi00}$^b$ and \citealt{ori03}). }
\label{tab_giant}
\begin{tabular}{@{}l cccccccccccc}
\hline
Species 	&\multicolumn{2}{c}{M\,13}&\multicolumn{2}{c}{$\omega$\,Cen}\\
		&Barnard\,29	&Giants&ROA\,5701&Giants\\
\hline \\
C     	&$<$6.64	&$\cdots$	&$<$6.54	&6.86	\\
N     	&7.45  		&$\cdots$	&7.06 		&$\cdots$   \\
O   	&7.50  		&7.53  		&7.76 		&7.45	   \\
Mg  	&6.10  		&6.14  		&6.15 		&6.26	   \\
Si  	&6.45  		&6.38  		&6.17 		&6.26	   \\
Al  	&5.14  		&5.75$^a$	&$<$4.87  	&4.87$^b$    \\
S	&5.72  		&$\cdots$	&5.62 		&$\cdots$   \\
Fe  	&6.07  		&5.90	  	&$<$6.02  	&5.92    \\
\hline
\end{tabular}
\end{center}
\end{table}

The absolute abundances of Barnard\,29 and ROA\,5701 are given in Table \ref{tab_giant}, along with those of giants from their respective clusters, namely M\,13 \citep{{coh05},{yon06}} and $\omega$\,Cen \citep{{smi00},{ori03}}. 
In general the abundances compare well, to within the errors. 
A notable exception is the Al abundance of Barnard\,29, which is lower than that of the giants. However, \citet{yon06} and \citet{sne04} have both shown that the aluminium abundances in the giants range from [Al/Fe] $\sim$ 0 -- 1.2 dex. The giants with [Al/Fe] $\sim$ 0.3 have similar O abundances to those found here, suggesting that the Barnard\,29 abundance compares well with a selection of the giant population.

\subsection{Evolutionary discussion}
\label{sec_evo}

\subsubsection{Dredge-up}

\citet{tho06} previously studied ROA\,5701, suggesting that the 
observed abundance patterns implied that the star had left the AGB prior to the onset of the 
third dredge-up, as have \citet{con94} to similarly explain the abundance patterns in Barnard\,29. 
Dredge-up occurs in stars during their AGB evolution, as products of hydrogen and helium burning are moved from the core to the surface, altering the surface composition \citep{ibe83}. Briefly, during the first dredge-up, the surface O sees little change while C decreases as N increases. The second dredge-up sees C and O mostly converted into N. At the third dredge-up, triple--$\alpha$ products are brought to the surface and C is enhanced relative to O and N. 

Both Barnard\,29 and ROA\,5701 exhibit abundance patterns compatible with no third dredge-up. The N abundances are strongly enhanced relative to both C and O, especially for Barnard\,29, while O is larger than C. This is consistent with the previous studies of both stars, with the exception of the lower N abundance of \citet{moe98} for ROA\,5701. Thus the abundances of CNO appear compatible with the suggestion that the third dredge-up may not have occurred.

The low C abundance could be explained in terms of hot bottom burning.  \citet{der05} found V435\,Oph to have a low metallicity and no enhancement of C, suggesting hot bottom burning as an explanation of the observed abundances. 
In hot bottom burning (HBB; \citealt{moo01}), C is converted to N at the base of the convective envelope, when temperatures become large enough for CNO processing. \citet{her05} indicates that HBB may have an important role at extremely low metallicities. 
However, \citet{dan96} suggest that Population {\sc i} stars of initial mass M $\geq$ 5 M$_{\sun}$, and Population {\sc ii} stars of M $\geq$ 3.5 M$_{\sun}$ experience HBB. As the initial mass of the stars is unknown, no conclusions can be made regarding this process.

\subsubsection{Gas-dust separation}

Although the lack of a third dredge-up appears a likely explanation for the abundance patterns of the light elements in 
both stars, \citet{moe98} and \citet{dix98} explain the abundances of the heavier elements
in terms of the gas-dust separation process. The basic premise involves grain formation that occurs in the circumstellar shells of stars during the AGB evolution. Dust particles are removed by radiation pressure and the `cleaned' gas is re-accreted onto the surface \citep{{mat92},{wat92}}. This results in abundances similar to the ISM, i.e. CNO and S near solar, while other elements such as Mg, Al, Si and Ca follow the low Fe abundance \citep{van95} and see for example the RV\,Tauri stars UY\,CMa, HP Lyr and BZ Sct \citep{gir05}.
\citet*{par92} found that, in the atmospheres of post-AGB stars, metals with a high condensation temperature, e.g. Fe, are more likely to condense into dust grains, compared to elements such as C, N, O and S with low condensation temperatures which stay in the gas phase with no significant depletions observed.  

\citet{moe98} suggested that this process occurred as they found a low Fe value, with N and O virtually undepleted, while \citet{dix98} also cite the low Fe abundance as evidence for the process. For this scenario, similar depletions of Mg, Si and Fe would therefore be expected. 
Examining the differential ISM abundances (Table \ref{tab_diff}) it is possible to determine if the stars follow the abundance trends seen in gas-dust separation. For Barnard\,29, we find differential abundances  of [Mg/H] = --1.3, [Al/H] = --1.2, [Si/H] = --0.8, [S/H] = --1.4 and [Fe/H](optical) = --1.4 (relative to the ISM). In addition, for ROA\,5701 the differential abundances are [Mg/H] = --1.3, [Si/H] = --1.1 and [S/H] = --1.5.  
The Mg, Al and Fe (optical) abundances of Barnard\,29 agree with the value for S and the metallicity and giant abundances of M\,13 ([Fe/H] = --1.6, \citealt{yon06}), although the Si abundance estimate is somewhat larger, possibly due to its sensitivity to the atmospheric parameters. 
For ROA\,5701, the S abundance agrees (to within the errors) with the Mg and Si values 
and is consistent with the metallicity and giants of $\omega$\,cen ([Fe/H] = --1.6, \citealt{ori03}). 
Similar results would be obtained if the abundances were compared with young B-type stars instead of the ISM. 

For gas-dust separation, CNO and S would be expected to be relatively undepleted and not in agreement with Mg, Si, Al or Fe. However, we find that the S abundance is generally in agreement with the Mg, Si, Al and Fe(optical) abundances, suggesting that gas-dust separation has not occurred. 
Also this process is generally confined to binary systems. The possible binarity of our two stars was investigated using radial velocity measurements. However a literature search produced sparse results, indicating no conclusions can be made. 
Additionally, the position of the stars on log~{\em g}/$T_{\rm eff}$ or log\,L/$T_{\rm eff}$ diagrams \citep{{con94},{blo95}} suggest core masses for both stars of $\sim$ 0.55 M$_{\sun}$, using log\,(L/L$_{\sun}$) = 3.2 \citep{lan92} for ROA\,5701 and log\,(L/L$_{\sun}$) = 3.23 \citep{ade94} for Barnard\,29. \citet{con93} found that stars of this average mass do not show an infrared excess, which is supported by the lack of {\sc IRAS} detection for both stars, and no emission features observed in H$\alpha$ for ROA\,5701. These results suggest there is no evidence of a circumstellar dust shell present for either star.	
Hence the gas-dust separation process appears unlikely to be responsible for the observed abundance patterns.

\subsubsection{UV lines}

In this paper, we find a difference of 0.7 dex and $<$1.0 dex for Barnard\,29 and ROA\,5701 respectively, between the optical and ultraviolet derived iron abundances. Analyses of the Fe\,{\sc iii} lines in other hot post-AGB stars, such as BD\,+33${^\circ}$2642 \citep*{nap94} and HD\,177566 \citep{{ken94a},{ken94b}}, show similar Fe abundance trends to those found here. 
\citet{nap94} analysed optical and {\it IUE} spectra, and their results led them to suggest the occurrence of gas-dust separation, as He was near solar, C, N, O, Mg and Si depleted by $\sim$1.0 dex, and Fe depleted by $\sim$2.0.  
\citet{ken94b} set upper limits for the Fe abundance from Fe\,{\sc iii}
lines in both optical and ultraviolet spectra, finding a difference of $\sim$0.5 dex between the two sets of data. 
Similarly, \citet{moe98} derived an Fe abundance of [Fe/H] = 6.95 dex from the Fe\,{\sc iii}
lines in the {\it IUE} spectra of the main-sequence B-type star $\gamma$\,Peg, compared to [Fe/H] = 7.56 dex from optical Fe\,{\sc iii} transitions.
In addition, \citet*{gri96} in their analysis of $\iota$\,Her found that the ultraviolet Fe abundance was $\geq$0.5 dex below the near solar value derived from optical lines.
\citet{gie92} and \citet{pro01} studied the B-type supergiant HD\,51309. While the former state two different iron abundance estimates depending on the temperature structure used, there is a difference of $\geq$0.75 dex found between their optical results (from Fe\,{\sc ii} and Fe\,{\sc iii} lines) and the {\it IUE} iron abundance of \citet{pro01}. 

Dufton (in preparation) have observed similar abundance discrepancies for young B-type stars in the 
Small and Large Magellanic Clouds (SMC, LMC), as well as in the Magellanic Bridge (MB). 
They determined Fe abundances from GHRS/HST and STIS/HST spectra, using the same methods and Fe\,{\sc iii}
transitions as described here, and found that these were consistently lower (more than 0.5 dex) than the 
estimated metallicities of the SMC, LMC and MB regions.

It is possible to derive the abundance of other elements using ultraviolet spectra, but that is not possible here as the GHRS data employed do not cover a large enough spectral range.  
However, {\it IUE} spectra have been used by a number of authors to derive the abundances of several light elements. 
\citet{con91} obtained a C abundance using both ultraviolet and optical spectra, finding the abundances to be compatible. \citet{ken94b} found similar differential abundances for C, N and Si between their optical and {\it IUE} spectra, with differences being less than 0.15 dex.  
Also, the Mg abundance obtained by \citet{nap94} showed a difference of 0.05 dex between their optical and ultraviolet values. 
However, the ultraviolet abundances obtained by \citet{nap94} for C and S are lower by 0.5 and 0.4 dex, respectively, than the optical abundances determined here (see Section \ref{sec_b29chem}).
Despite this result, the use of ultraviolet spectra to determine the abundances of light elements generally appears to be reliable and do not suffer the same problem observed with iron. 

\subsubsection{Oscillator strengths}

The above studies, in conjunction with the present work, strongly indicate that 
the ultraviolet Fe\,{\sc iii} lines yield systematically low Fe abundances, possibly as a 
result of errors in the adopted atomic data, in particular f-values.
Currently, atomic data from Kurucz models (http://nova.astro.umd.edu, \citealt{{hub88},{hub95},{hub98}}) are used in the {\sc tlusty} grids. 
We have therefore performed a literature search of f-values for Fe\,{\sc iii},
to investigate if more recent calculations or measurements would produce different Fe abundances. 
Table \ref{tab_os} lists the six strongest transitions observed in the GHRS/HST 
spectra, along with the oscillator strengths from Kurucz, and the more recent results
of \citet{nah96} and \citet{ton05}. 
There are little differences between the sets of atomic data, so these will not solve the
Fe abundance problem. However we note that further calculations and measurements for these
Fe\,{\sc iii} transitions are planned in the future. 

Despite numerous studies of the ultraviolet Fe\,{\sc iii} transitions, the line list for the spectral region employed in this paper may not be complete. However, it is assumed that, if this is the case, any lines excluded during the fitting procedure would be compensated by making other transitions too strong. This would potentially lead to an over estimate in the derived abundance, which would not solve the abundance discrepancy observed.

\begin{table*}
\begin{minipage}{\textwidth}
\begin{center}
\caption{Strongest transitions of Fe\,{\sc iii} seen in the ultraviolet spectra of Barnard\,29 and ROA\,5701. Oscillator strengths are taken from Kurucz, \citet{nah96}$^a$ and \citet{ton05}.}
\label{tab_os}
\begin{tabular}{cccccccccccccc}
\hline
Transition$^a$			& g$_i$$^a$ & g$_f$$^a$ &Wavelength 	 & \multicolumn{3}{c}{log~{\em gf} values}     \\
	  		  	&	    &	 	&(\AA) 	   	 & Kurucz& \citet{nah96} & \citet{ton05} \\
\hline	\\
a${}^5$G$^e$--z${}^5$F$^o$	& 5	& 5	 & 1869.841	 & -0.666	&  -0.672 &$\cdots$	  \\
a${}^5$G$^e$--z${}^5$F$^o$	& 5	& 3	 & 1871.152	 & -0.121	&  -0.116 &$\cdots$	  \\
b${}^5$D$^e$--y${}^5$D$^o$	& 9	& 9	 & 1877.990	 &  0.238	&   0.273 &$\cdots$	  \\
a${}^5$G$^e$--z${}^5$F$^o$	& 13	& 11	 & 1890.664	 &  0.434	&   0.407 &$\cdots$	  \\
a${}^7$S$^e$--z${}^7$P$^o$	& 7	& 9	 & 1895.456	 &  0.461	&   0.415 & 0.436	  \\
a${}^3$I$^e$--y${}^3$H$^o$	& 13	& 11	 & 1896.814	 &  0.480	&   0.398 &$\cdots$	  \\

\hline
\end{tabular}
\end{center}
\end{minipage}
\end{table*}

\subsubsection{Further work}

To investigate the Fe abundance discrepancy 
further will require very high S/N data for Barnard\,29 and ROA\,5701,
sufficient to ascertain the reliability of the observed optical Fe\,{\sc iii} lines and observe further Fe\,{\sc iii} lines in both stars. Unlike the ultraviolet transitions, the optical
lines are sufficiently weak, even in main-sequence Population\,{\sc i} B-type stars,
for a detailed line-by-line differential abundance analysis to be performed. This will 
hence allow the removal of systematic effects such as the dependence on f-value, and hence the determination
of accurate Fe abundances, for comparison with the ultraviolet measurements. 
We plan such observations in the near future.

\section{Conclusions}

The abundance patterns observed for both Barnard\,29 and ROA\,5701 indicate that the stars
have not undergone the third dredge-up. 
In addition, the gas-dust separation process suggested by previous authors appears unlikely. 
The optical spectra imply abundances for the Fe and other elements, which are compatible with those found in the late-type giants, 
while the Fe\,{\sc iii} lines in the ultraviolet produce lower abundances than expected, 
as found in several other studies. 
This strongly suggests that the ultraviolet Fe\,{\sc iii} do not provide reliable abundance indicators.
Future studies require high S/N optical data, in order to obtain accurate Fe abundance estimates.
A re-evaluation of the Fe\,{\sc iii} f-values, either theoretically or experimentally, would
also be highly desirable.

\section*{acknowledgments}

Some of the data presented in this paper were obtained from the Multimission Archive at the Space Telescope Science Institute (MAST). STScI is operated by the Association of Universities for Research in Astronomy, Inc., under NASA contract NAS5-26555. Support for MAST for non-HST data is provided by the NASA Office of Space Science via grant NAG5-7584 and by other grants and contracts.
{\sc iraf} is distributed by the National Optical Astronomy Observatories, U.S.A.
We would like to thank Ian Howarth for his continued help. 
HMAT acknowledges financial support from the Northern Ireland Department of Education and Learning (DEL). 
FPK is grateful to AWE Aldermaston for the award of a William Penney Fellowship. 
JVS thanks the staff of the McDonald observatory for their expert assistance in taking these observations and the Particle Physics and Astronomy Council for financial support. 
DLL thanks the Robert A. Welch Foundation of Houston, Texas for their support.
AAZ thanks the SAAO for hospitality during a sabbatical visit.

{}

\label{lastpage}

\end{document}